\documentclass[aps,pra,twocolumn,showpacs,superscriptaddress,10pt,letterpaper,nofootinbib,showkeys]{revtex4-1}

\usepackage{graphicx}% Include figure files
\usepackage{dcolumn}% Align table columns on decimal point
\usepackage{bm}% bold math
\usepackage{color}% add textcolor
\usepackage{amsmath}% for subequations environment
\usepackage{mdframed}
\usepackage{physymb}
\usepackage{enumitem}

\begin{document}

\title{Computation across the curriculum: What skills are needed?}

\author{Marcos D. \surname{Caballero}}
   \affiliation{Department of Physics and Astronomy \& CREATE for STEM Institute, Michigan State University, East Lansing, MI 48824}
% \author{Laura \surname{Merner}}
%    \affiliation{American Institute of Physics, College Park, MD 20740}
%  \author{Robert C. \surname{Hilborn}}
%    \affiliation{American Association of Physics Teachers, College Park, MD 20740}
%  \author{Norman J. \surname{Chonacky}}
%    \affiliation{Department of Applied Physics, Yale University, New Haven, CT 06511}

%\date{\today}

\begin{abstract}
Computation, the use of a computer to solve, simulate, or visualize a physical problem, has revolutionized how physics research is done. Computation is used widely to model systems, to simulate experiments, and to analyze data. Yet, in most undergraduate programs, students have little formal opportunity to engage with computation and, thus, are left to their own to develop their computational expertise. As part of a larger project to study how computation is incorporated in some undergraduate physics programs (and how it might be incorporated further), we convened a mini-conference and conducted a series of interviews with industry professionals, academic faculty, and employed bachelor's graduates who make use of computation in their everyday work. We present preliminary results that speak to how participants developed the requisite skills to do professional computational work and what skills they perceive are necessary to conduct such work.
\end{abstract}

\maketitle

\section{Introduction\label{sec:intro}}

While computation has revolutionized modern science research \cite{wolfram2002new}, much of the undergraduate physics curriculum remains as it was in the 1960s \cite{meltzer2015brief} -- e.g., emphasizing the construction of analytical solutions to long-solved problems. With advent of powerful computers as well as the development of high-level programming languages, we are capable of modeling phenomenon that a generation ago would have required sophisticated tools. And yet, many of our programs have not yet incorporated computation into the undergraduate experience \cite{Chonacky:2008gq}.

A new push to incorporate computation into the undergraduate curriculum is occurring \cite{McIntyre:2008ej,Chabay:2008jw,Caballero:2012hu,HjorthJensen:vj} that builds on the successes of the past \cite{diSessa:1986uz,1993PhyEd..28..102S}, but also aims to incorporate results from physics education research \cite{Caballero:2014gn}. As a step towards building this effort, we convened a two-day, 30-person conference of industry professionals, academic faculty, and employed physics bachelor's graduates who make use of computation in their everyday work. The overarching purpose of this conference was to articulate goals for a national survey that will investigate the state of computation in undergraduate physics departments. Follow-up interviews were conducted with a purposefully selected group of participants to further develop our understanding of what they perceive as goals for teaching computation, how their own computational learning was supported, and how their current organization continues to support their (and others') learning.

In this paper, we report on the skills physics bachelor's graduates should develop from the perspective of conference and interview participants. This study makes use of field notes and interviews, and analyzes this data from a phenomenongraphic perspective to represent diversity of participants' views while seeking coherence among those views \cite{marton1981phenomenography}. This preliminary study acts as a starting point for developing more concrete curricular and pedagogical structures for teaching computation in physics courses. %The remainder of the paper is organized as follows: In Sec.~\ref{sec:context}, we discuss briefly the goals of our study and how we are looking at our data to achieve those goals. We present an overview of the data and how it was collected in Sec.~\ref{sec:data}. Because of space restrictions, we avoid providing too much detail on the analysis of data and rather present the results of this analysis including supporting data in Sec.~\ref{sec:results}. Finally, we discuss nuances of the views presented (Sec.~\ref{sec:results}) and implications for future work in Sec.~\ref{sec:discussion}.

\section{Data}\label{sec:data}

The discussions at the conference and during the post-conference interviews provided information on a wide range of issues around computation. The structure of the conference was such that we could collect field notes of small group and larger plenary discussions. Researchers acted as participant-observers who facilitated conversation, but allowed participants to carry the conversation in both small group and larger discussions. While all researchers collected their own notes, a single researcher was dedicated to collecting field notes. Field notes were discussed and edited during and after the conference to represent each researcher's recollection of the discussions.

During the conference, we identified participants with whom to conduct follow-up interviews. We gave precedence to employed physics bachelor's graduates and those who employ them including academic faculty and industry professionals. 
%The pool of possible interviewees was further winnowed through interactions during the conference -- selecting participants who presented clearly-articulated views and thoughts about computation in the undergraduate curriculum. 
Ultimately, we interviewed one-fifth of conference attendees. One-hour interviews were conducted with two employed physics bachelor's graduates (Britta \& Annie), two academic faculty who advise doctoral candidates on computational physics research (Frankie \& Jeff), one industry professional (Pierce), and one academic faculty who develops instructional materials for physics courses (Leonard). Some similarities, but also clear differences between the experiences of junior and senior interviewees emerged from the analysis.

We offer an overview of the analysis of our data (Sec.~\ref{sec:context}) to give sufficient space to present our results (Sec.~\ref{sec:results}) and discussion (Sec.~\ref{sec:discussion}).

\section{Assumptions and Methods}\label{sec:context}

The data that we collected have provided a wealth of information about many aspects of computation in general, computational instruction, and the necessary skills for undergraduates to develop. Our analysis in this paper has focused exclusively on participants' perceptions of the skills that undergraduates need to develop with regard to computation. We have not wholly neglected these other important aspects of our data, but for the purposes of this paper we have foregrounded participants' discussions around computational skills in our analysis.

%Further, we acknowledge that our active participation in the conference may have colored the data and our interpretation of it. A particular line of discussion may have been promoted or shut down by facilitators and bids to change topic may have been taken up or dismissed depending on the facilitator's perception of interest or timing. However, the coherence across discussions and interviews represented in our data suggests that the main results are likely stable in the population of attendees (Sec.~\ref{sec:results}), but there might be nuances to these main results that are not well-represented in this paper.

We should also note that our participants and interviewees were not (generally) grounded in educational research and do not appear to subscribe to any particular definition of ``skill.'' Thus, we have taken a broad view of ``skills,'' as it appears our participants and interviewees did, which includes knowledge of certain concepts, specific practices, and ways of thinking.

To analyze the collected data, we open-coded field notes and interviews under the aforementioned assumptions (i.e., looking for specific discussions of computational skills). This initial coding process revealed a number of similarities among the conference participants' discussions and interviewee's comments. Further review of tagged interview sections fleshed out these commonalities as well as the nuance between different interviewee's perceptions. In what follows, we present our interpretation of the common computational skills discussed by conference participants and interviewees along with some of the observed nuances.

\section{Computational Skills}\label{sec:results}

Conference and interview participants presented a number of different computational skills in discussion and in interviews. Our categorization of these skills is based on a triangulated coding of field notes and interview responses. We do not claim that ours is the only categorization possible for such data, but that it is consistent with our assumptions and that it lends itself to be interpretable for our purposes readily. We discuss each category of skills beginning with those articulated in the conference, offer sample responses from interviews that provide evidence for the presence of each category, and comment on the nuance between interview participant's responses around each category.

\begin{center}
{\bf Knowledge of specific physics \& computational concepts}
\end{center}

Central to much of the early discussion at the conference was the particular physical and computational concepts that were necessary for students to learn during their degree program. The physics topics that were discussed mirrored the canon of topics presented in a typical physics major. The computational concepts that were raised spanned a variety of iterative concepts and included well-established numerical methods (e.g., relaxation \& Monte Carlo methods) as well newer concepts for data science (e.g., anomaly detection, clustering).

The centrality of specific physical and computational concepts was present to a lesser degree in interviews (by design), but was still articulated by all interviewees. The specific topics raised depended directly on the interviewee's line of work or research. For example, Britta, a first-year graduate student in soft matter, responded to a line of inquiry about what a new graduate student might need to know to work in her lab,

\begin{quote}
\noindent {\bf Britta} {\it...a good, solid classical mechanics background as well. We're having trouble implementing rigid body dynamics in our code now because we're not sure if there's frictional forces going on with our particle or it's the lubrication force that is really dominating. So if you don't know the physics going on behind that, how can you put it into the computer?}
\end{quote}

In this snippet, Britta mentions a number of physics concepts that are needed in her work (e.g., rigid body dynamics) explicitly, but she also mentions putting that physics into the computer, which, from other parts of her interview, point to different forms of numerical integration. More senior interviewees (Jeff, Pierce, Leonard, and Frankie) also discussed specific physics and computational concepts used in their work, but they did not often couch those discussions in particular projects as Britta does. Instead, it was common for senior interviewees to describe broader aspects of learning computing.

\begin{center}
{\bf Use of specific computational practices}
\end{center}

In addition to students knowing particular concepts, they must engage with specific computational practices. Conference attendees raised a number of practices that can be categorized broadly as planning, modeling, programming, debugging, analysis, and visualization. These practices are particular kinds of actions that attendees felt were important for students to develop fluency with. Interviewees echoed these practices and further defined them. For example, Britta discussed the importance of planning (i.e., writing pseudocode),

\begin{quote}
\noindent {\bf Britta} {\it Sometimes I would have an idea and I would think about the physics but I never really articulate it. Writing pseudocode and working through the mathematics before vomiting the code out that made a difference.}
\end{quote}

Remarks from interviewees suggested that these practices are important for students to engage in modern science work (all), but are often lacking from the undergraduate experience (Jeff, Leonard, and Frankie).

\begin{center}
{\bf Computational ways of thinking}
\end{center}

During the conference, attendees often alluded to ``computational thinking'' or ``thinking iteratively '' as desirable skills for students to develop. Such terms appear in the literature, but are often nebulous \cite{wing2006computational}. In interviews, some more senior interviewees (Jeff, Pierce, and Leonard)  raised the issue of computational thinking spontaneously and unpacked it to various degrees. For example, Jeff, a professor who conducts computational astrophysics research, discussed computational thinking as breaking down a problem into its finest components. He, then, mentioned how difficult that is to learn,

\begin{quote}
\noindent {\bf Jeff} {\it [thinking computationally is] thinking about it [the task/problem] in the very specific steps of what needs to be done that often is hard for people to get into...once you learn to break it down...I think getting to think that way, that's the challenging part.}
\end{quote}

Pierce and Leonard also raised issues associated with breaking down a complex task into bits of code, but Pierce and Jeff went further to discuss how important thinking discretely (e.g., thinking about the accumulation of small chunks) was to the computational endeavor. More junior interviewees (Annie and Britta) did not raise issues of computational thinking specifically. Some of their discussions about specific tasks they completed could have encapsulated computational thinking, but direct probes were not taken up. It might not be surprising that more junior interviewees struggled to articulate ideas around computational thinking as they have not had the benefit of years or decades of working with computation that more senior interviewees have had.

% \begin{quote}
% \noindent {\bf Interviewer} {\it ...were there particular ways of thinking that you think were developed because you were doing computing that helped you do that [a specific task]?}
% \noindent {\bf Annie} {\it ...maybe logic, different logic, different syntaxes. I guess logic might be the word for it.}
% \end{quote}

\begin{center}
{\bf Ways of connecting mathematics, physics, and computational ideas}
\end{center}

In their discussions, conference participants extended these ways of thinking beyond ``iterative thinking'' to include what we will refer to as the {\it math-physics-computing connection}. According to participants, students who have developed this connection are able to move fluidly between mathematical, physical, and computational ideas and are able to see how each affects the other. In interviews, each participant discussed the math-physics-computing connection. Pierce, an engineer who manages a team that is developing hardware and software for exascale supercomputing, mentions how challenging it is to find such expertise,

\begin{quote}
\noindent {\bf Pierce} {\it ...one of the things that's very scarce in the people who we look for is the knowledge of the actual physics and the mathematics and the numerical mathematics that goes into a lot of the simulation programs.}
\end{quote}

It might appear that Pierce's discussion of knowledge is simply ideas or concepts related to each of these different areas, but it's more clear from other parts of his interview that he is discussing connections,

\begin{quote}
\noindent {\bf Pierce} {\it If you don't really have a fundamental understanding of what you are doing in the mathematics that will affect the representation of the physics, then you could be doing things [with the computer] that are just going to be useless later on.}
\end{quote}

While all of the more senior interviewees raised the connection between mathematics, physics, and computing explicitly, more junior interviewees (Britta) discussed them more implicitly -- often couched in the particulars of their work. This is, again, not terribly surprising given the much wider experience with computation that more senior interviewees have had.

\begin{center}
{\bf Meta-knowledge about computation}
\end{center}

In addition to knowledge and practices that support doing computation, conference participants voiced that students should develop meta-knowledge about what computing can do and what tools are appropriate for certain tasks. In interviews, developing this meta-knowledge was presented by more senior interviewees (Frankie, Jeff, and Leonard), but discussions about choosing specific tools were absent. Meta-knowledge discussions were absent in interviews with more junior interviewees. Frankie, a professor who runs a soft matter lab, discussed how students should develop a meta-level understanding about what computation can buy them.

\begin{quote}
\noindent {\bf Frankie} {\it I want students to recognize that when [a diffusive system] runs to equilibrium, you're solving the Laplace equation because the time derivative is zero when in equilibrium.}
\end{quote}

Here, Frankie is speaking about helping students make the connection between solving the diffusion equation numerically and solving Laplace's equation analytically.
%The analytic solution is typically an infinite series.
She later articulates why she wants students to do this,

\begin{quote}
\noindent {\bf Frankie} {\it I had students solve problems numerically that they could also solve analytically...calculate the first 50 terms [of an infinite series solution to Laplace's equation] and compare that to the computational [solution]. Once you gain the ability to get that more or less right, then you can start solving problems where you don't know the answer.}
\end{quote}

Frankie is both helping students understand analytic solutions that contain infinite series and demonstrating that the computer can produce an accurate solution that fits with students' analytical work. We infer that Frankie is, in part, working to help students build meta-knowledge about what computation can do. Other senior interviewees expressed similar sentiments, for example, Jeff and Leonard suggested that students should understand that most problems cannot be solved analytically.

\begin{center}
{\bf Developing professional practices}
\end{center}

Finally, conference participants raised issues of professional practice. We distinguish these practices from computational practices because they are more closely related to working in a team environment (e.g., co-developing software). For many participants, documenting code and using version control were paramount to working well on teams that develop complex codes. These sentiments were echoed by Jeff, who develops an open-source package for astrophysical calculations and Britta, who is working from poorly documented code.

\begin{quote}
\noindent {\bf Jeff} {\it ...when you write code in a research group, you are going to be forced to support it until you die...even though it takes 20 percent longer, it's a good idea to make it readable and document it and use version control...}
\end{quote}

\begin{quote}
\noindent {\bf Britta} {\it ...we have a code that was written by a previous postdoc and he does not like to respond to emails. So, there's a lot of error messages and I have to try to figure out why.}
\end{quote}

In other parts of her interview, Britta expressed frustration with her project (e.g., persistent errors). She did not discuss how her frustration might be connected to the professional practices that Jeff mentioned (e.g, properly documented code). As Britta has worked in computational physics for 4 years and exclusively in small research groups, it is likely that she has not yet recognized the importance of these professional practices. By contrast, Jeff's years of experience developing software has required him to make use of these practices extensively.

\section{Discussion and Conclusions}\label{sec:discussion}

In this study, we collected diverse ideas presented by conference and interview participants into several preliminary categories of computational ``skills.'' As discussed (Sec.~\ref{sec:context}), we have taken a broad view of computational skills. The categories we have presented (Sec.~\ref{sec:results}) do not represent an exhaustive list of computational skills, but this list does collect the different ways that our population spoke about what computational skills students should develop in their undergraduate physics program.

It is worth noting that these categories are not without nuance. For example, we have found a recurring divide among the ways in which different generations of computational physicists discuss what skills are important for students to learn. More junior interviewees were more likely to ground their discussion in particular projects. Additionally, some skills (e.g., computational ways of thinking) were generally absent from interviews with more junior interviewees.
Both of these results likely stem from the fact that more senior interviewees have more and broader experiences with computation. These more senior participants appear to draw on and synthesize those experiences in our interviews.
%Additional divides appear with academic faculty and other interviewees (i.e., meta-knowledge about computation) as well as those who work in large collaborations and those who do not (i.e., developing professional practices).

The result of this research is a preliminary list of computational skills that can act as starting point for developing more concrete curricular and pedagogical structures for teaching computation. A recurring theme not presented in this study was that all interview participants developed their computational physics expertise outside of the physics curriculum; they were self-taught. It was the sentiment of conference participants and interviewees that this self-teaching practice is insufficient to deal with the needs placed on future physicists by graduate schools and industry. That is, we require new and different ways of incorporating computational instruction going forward.

\begin{acknowledgments}
We thank Laura Merner, Norman Chonacky, and Bob Hilborn for their work at the conference. We also thank the members of PERL@MSU for their useful comments at various stages of this work.  This work is supported the National Science Foundation (DUE \#1431776).
\end{acknowledgments}

\bibliography{perc2015_comp}

\begin{thebibliography}{12}
\expandafter\ifx\csname natexlab\endcsname\relax\def\natexlab#1{#1}\fi
\expandafter\ifx\csname bibnamefont\endcsname\relax
  \def\bibnamefont#1{#1}\fi
\expandafter\ifx\csname bibfnamefont\endcsname\relax
  \def\bibfnamefont#1{#1}\fi
\expandafter\ifx\csname citenamefont\endcsname\relax
  \def\citenamefont#1{#1}\fi
\expandafter\ifx\csname url\endcsname\relax
  \def\url#1{\texttt{#1}}\fi
\expandafter\ifx\csname urlprefix\endcsname\relax\def\urlprefix{URL }\fi
\providecommand{\bibinfo}[2]{#2}
\providecommand{\eprint}[2][]{\url{#2}}

\bibitem[{\citenamefont{Wolfram}(2002)}]{wolfram2002new}
\bibinfo{author}{\bibfnamefont{S.}~\bibnamefont{Wolfram}},
  \emph{\bibinfo{title}{{A New Kind of Science?}}}
  (\bibinfo{publisher}{{Wolfram Media}}, \bibinfo{year}{2002}).

\bibitem[{\citenamefont{Meltzer and Otero}(2015)}]{meltzer2015brief}
\bibinfo{author}{\bibfnamefont{D.~E.} \bibnamefont{Meltzer}} \bibnamefont{and}
  \bibinfo{author}{\bibfnamefont{V.~K.} \bibnamefont{Otero}},
  \emph{\bibinfo{title}{A brief history of physics education in the united
  states}}, \bibinfo{journal}{Am. J. Phys.} \textbf{\bibinfo{volume}{83}},
  \bibinfo{pages}{447} (\bibinfo{year}{2015}).

\bibitem[{\citenamefont{Chonacky and Winch}(2008)}]{Chonacky:2008gq}
\bibinfo{author}{\bibfnamefont{N.}~\bibnamefont{Chonacky}} \bibnamefont{and}
  \bibinfo{author}{\bibfnamefont{D.}~\bibnamefont{Winch}},
  \emph{\bibinfo{title}{{Integrating computation into the undergraduate
  curriculum: A vision and guidelines for future developments}}},
  \bibinfo{journal}{Am. J. Phys.} \textbf{\bibinfo{volume}{76}},
  \bibinfo{pages}{327} (\bibinfo{year}{2008}).

\bibitem[{\citenamefont{McIntyre et~al.}(2008)\citenamefont{McIntyre, Tate, and
  Manogue}}]{McIntyre:2008ej}
\bibinfo{author}{\bibfnamefont{D.~H.} \bibnamefont{McIntyre}},
  \bibinfo{author}{\bibfnamefont{J.}~\bibnamefont{Tate}}, \bibnamefont{and}
  \bibinfo{author}{\bibfnamefont{C.~A.} \bibnamefont{Manogue}},
  \emph{\bibinfo{title}{{Integrating computational activities into the
  upper-level Paradigms in Physics curriculum at Oregon State University}}},
  \bibinfo{journal}{Am. J. Phys.} \textbf{\bibinfo{volume}{76}},
  \bibinfo{pages}{340} (\bibinfo{year}{2008}).

\bibitem[{\citenamefont{Chabay and Sherwood}(2008)}]{Chabay:2008jw}
\bibinfo{author}{\bibfnamefont{R.}~\bibnamefont{Chabay}} \bibnamefont{and}
  \bibinfo{author}{\bibfnamefont{B.}~\bibnamefont{Sherwood}},
  \emph{\bibinfo{title}{{Computational physics in the introductory
  calculus-based course}}}, \bibinfo{journal}{Am. J. Phys.}
  \textbf{\bibinfo{volume}{76}}, \bibinfo{pages}{307} (\bibinfo{year}{2008}).

\bibitem[{\citenamefont{Caballero et~al.}(2012)\citenamefont{Caballero,
  Kohlmyer, and Schatz}}]{Caballero:2012hu}
\bibinfo{author}{\bibfnamefont{M.}~\bibnamefont{Caballero}},
  \bibinfo{author}{\bibfnamefont{M.}~\bibnamefont{Kohlmyer}}, \bibnamefont{and}
  \bibinfo{author}{\bibfnamefont{M.}~\bibnamefont{Schatz}},
  \emph{\bibinfo{title}{{Implementing and assessing computational modeling in
  introductory mechanics}}}, \bibinfo{journal}{Phys. Rev. ST - PER}
  \textbf{\bibinfo{volume}{8}}, \bibinfo{pages}{020106} (\bibinfo{year}{2012}).

\bibitem[{\citenamefont{Hjorth-Jensen et~al.}(2008)\citenamefont{Hjorth-Jensen,
  M{\o}rken, Myhre, and S{\o}lna}}]{HjorthJensen:vj}
\bibinfo{author}{\bibfnamefont{M.}~\bibnamefont{Hjorth-Jensen}},
  \bibinfo{author}{\bibfnamefont{K.}~\bibnamefont{M{\o}rken}},
  \bibinfo{author}{\bibfnamefont{A.}~\bibnamefont{Myhre}}, \bibnamefont{and}
  \bibinfo{author}{\bibfnamefont{H.}~\bibnamefont{S{\o}lna}}, in
  \emph{\bibinfo{booktitle}{{Ripples -- Five Years of Flexible Learning at the
  University of Oslo}}} (\bibinfo{publisher}{University of Oslo},
  \bibinfo{year}{2008}).

\bibitem[{\citenamefont{diSessa and Abelson}(1986)}]{diSessa:1986uz}
\bibinfo{author}{\bibfnamefont{A.}~\bibnamefont{diSessa}} \bibnamefont{and}
  \bibinfo{author}{\bibfnamefont{H.}~\bibnamefont{Abelson}},
  \emph{\bibinfo{title}{{Boxer: a reconstructible computational medium}}},
  \bibinfo{journal}{Commun. ACM} \textbf{\bibinfo{volume}{29}},
  \bibinfo{pages}{859} (\bibinfo{year}{1986}).

\bibitem[{\citenamefont{Schecker}(1993)}]{1993PhyEd..28..102S}
\bibinfo{author}{\bibfnamefont{H.}~\bibnamefont{Schecker}},
  \emph{\bibinfo{title}{{Learning physics by making models}}},
  \bibinfo{journal}{Phys. Ed.} \textbf{\bibinfo{volume}{28}},
  \bibinfo{pages}{102} (\bibinfo{year}{1993}).

\bibitem[{\citenamefont{Caballero and Pollock}(2014)}]{Caballero:2014gn}
\bibinfo{author}{\bibfnamefont{M.~D.} \bibnamefont{Caballero}}
  \bibnamefont{and} \bibinfo{author}{\bibfnamefont{S.~J.}
  \bibnamefont{Pollock}}, \emph{\bibinfo{title}{{A model for incorporating
  computation without changing the course: An example from middle-division
  classical mechanics}}}, \bibinfo{journal}{Am. J. Phys.}
  \textbf{\bibinfo{volume}{82}}, \bibinfo{pages}{231} (\bibinfo{year}{2014}).

\bibitem[{\citenamefont{Marton}(1981)}]{marton1981phenomenography}
\bibinfo{author}{\bibfnamefont{F.}~\bibnamefont{Marton}},
  \emph{\bibinfo{title}{Phenomenography -- describing conceptions of the world
  around us}}, \bibinfo{journal}{Instr. Sci.} \textbf{\bibinfo{volume}{10}},
  \bibinfo{pages}{177} (\bibinfo{year}{1981}).

\bibitem[{\citenamefont{Wing}(2006)}]{wing2006computational}
\bibinfo{author}{\bibfnamefont{J.~M.} \bibnamefont{Wing}},
  \emph{\bibinfo{title}{Computational thinking}},
  \bibinfo{journal}{Communications of the ACM} \textbf{\bibinfo{volume}{49}},
  \bibinfo{pages}{33} (\bibinfo{year}{2006}).

\end{thebibliography}
\bibliographystyle{apsper}

\end{document}